\documentclass[12pt]{article}
\usepackage{amsmath}
\usepackage{amssymb}
\usepackage{bbm}
\usepackage{epsfig}
\usepackage{yfonts}
\newcommand{\be}{\begin{equation}}
\newcommand{\ee}{\end{equation}}
\newcommand{\bea}{\begin{eqnarray}}
\newcommand{\eea}{\end{eqnarray}}
\newcommand{\ba}{\begin{eqnarray}}
\newcommand{\ea}{\end{eqnarray}}

\newcommand{\beq}{\begin{equation}}
\newcommand{\eeq}{\end{equation}}
\newcommand{\beqa}{\begin{eqnarray}}
\newcommand{\eeqa}{\end{eqnarray}}
\newcommand{\beqar}{\begin{eqnarray*}}
\newcommand{\eeqar}{\end{eqnarray*}}

\newcommand{\reef}[1]{(\ref{#1})}

\newcommand{\ie}{{\it i.e.,}\ }
\newcommand{\comment}[1]{{\bf [[[#1]]]}}

\newcommand{\mt}[1]{\textrm{\tiny #1}}

\newcommand{\lgb}{\lambda_{\text{\tiny{GB}}}}

\newcommand{\nc}{N_c}

\newcommand{\la}{\lambda}
\newcommand{\lp}{\ell_{\mt P}}

\begin{document}

\title{Quantum corrections to screening  at strong coupling}
\smallskip\smallskip\smallskip
\author{{ Ajay Singh}$^{1,3}$\thanks{email: asingh@perimeterinstitute.ca} $~~$ and
{ Aninda Sinha}$^{2}$\thanks{email: asinha@cts.iisc.ernet.in}
\\[8mm]
\it $^1$Perimeter Institute for Theoretical Physics, \\
\it Waterloo, Ontario N2L 2Y5\\[2mm]
\it $^2$Centre for High Energy Physics, Indian Institute of Science,\\
\it C. V. Raman Avenue, Bangalore 560012, India \\ [2mm]
\it $^3$Department of Physics \& Astronomy and Guelph-Waterloo Physics Institute,\\
\it University of Waterloo, Waterloo, Ontario N2L 3G1, Canada}

\maketitle
\vskip 2cm


\begin{abstract}
We compute 
a certain class of corrections to (specific) screening lengths 
in strongly coupled nonabelian plasmas using the AdS/CFT correspondence.
In this holographic framework, these corrections arise from various
higher curvature interactions modifying the leading Einstein gravity
action. The changes in the screening lengths are perturbative in
inverse powers of the 't Hooft coupling or of the number of colours, as
can be made precise in the context where the dual gauge theory is
superconformal. We also compare the results of these holographic
calculations to lattice results for the analogous screening lengths in
QCD. In particular, we apply these results within the program of making
quantitative comparisons between the strongly coupled quark-gluon
plasma and holographic descriptions of conformal field theory.
\end{abstract}

\maketitle
\tableofcontents

\section{Introduction}
The AdS/CFT correspondence  \cite{Maldacena1,Aharony, Witten1, Gubser} has proved to be a useful tool in probing strongly coupled physics. 
The quark gluon plasma  \cite{shuryak} formed in heavy ion collisions in the RHIC and LHC may be strongly coupled and it is a useful exercise to see which 
features can be reproduced using the correspondence  \cite{cvj, talks}. One of the main reasons for this possible connection is the observation that a wide class of holographic theories \cite{Kovtun, buchelliu} have a small ratio of shear viscosity ($\eta$) to entropy density ($s$) \footnote{It has become a whole new industry to see which fluid in nature has the lowest $\eta/s$ \cite{lowest}.}, which is what appears to be needed to explain the RHIC and LHC data \cite{heinz}.

It is now known that while holographic theories describing isotropic plasmas\footnote{It has been recently claimed that anisotropy can lead to non-universal $\eta/s$ \cite{anis}.} using Einstein gravity have $\eta/s=1/4\pi$, this is no longer true when higher curvature corrections are taken into account \cite{Buchel,buchelbs,quantum}. While most of the work in the literature has focused on higher curvature corrections to transport coefficients, in this paper we will turn our attention to corrections to screening lengths originally considered in \cite{Bak1}. The screening lengths are defined through spatial correlators of Polyakov loops. When the spatial separation of the loops is large, the fall off of the connected contribution is exponential in separation, namely $e^{-|{\bf x}|/\xi}$, where $\xi$ is the screening length. The mass corresponding to the longest screening length is called the mass gap $m_{gap}$. On the holographic side, this corresponds to the lightest supergravity mode \cite{Csaki,Minahan,Constable,Brower} that is exchanged between two strings stretching between the boundary and the horizon. In Einstein gravity, this happens to be the time-time component of the metric. The mass corresponding to the exchange of the axion field is the Debye mass. Depending on which supergravity mode is exchanged, there is a corresponding screening length. Study of  pure gluon theories reveal little dependence on $\nc$ so it may be hoped that the differences between the QCD results and the large $\nc$ results are going to be small \cite{Teper1,hart}. The holographic screening masses actually work out to be larger than the lattice results \cite{Bak1} and it could be hoped that finite $\lambda, 1/\nc$ corrections would lead to a better agreement.

The most general higher curvature corrections at leading order are not known in IIB string theory. Only certain special classes of corrections have been worked out \cite{Gross, greenstahn, polipuri}. In the large $\la$ and large $\nc$ limit, it will turn out that the known curvature corrections are sufficient to compute the leading perturbative correction to certain screening lengths, namely the mass gap and the screening mass corresponding to a vector and a spin-2 exchange as explained below.  In certain
supersymmetric theories where one has AdS$_5\times M_5$ where $M_5$ is
smooth compact manifold, the leading higher curvature correction is $R^4$  and gives $1/\lambda^{3/2}$ and $\lambda^{1/2}/\nc^2$
corrections \cite{quantum,Buchel3}. In certain instances where the dual includes fundamental
fields, the leading corrections are $1/\nc$ coming from $R^2$ terms \cite{kats,Buchel1}. As we will argue below, knowledge of the higher curvature terms is sufficient to compute the leading corrections to the screening masses corresponding to the exchange of a graviton. Corrections to screening masses corresponding to the dilaton, axion (Debye mass), $C_2$, $B_2$ are all beyond the scope of current technology (see the discussion in \cite{Bak1}).

In a  broader context, the higher derivative modifications of the
leading supergravity action will expand the universality class of the
dual CFT, by modifying the parameters in the n-point functions of the
CFT. In particular, $R^3$ is a natural term that would appear but only
in a non-supersymmetric context \cite{Hofman}. This has been studied in recent toy models \cite{quasi1} where the coupling of the $R^3$ terms were not restricted to be small. In the context of the present work, we will work in a perturbative framework.  A further motivation in considering such corrections is to extend the phenomenology program initiated in \cite{Buchel1} and studied further in \cite{Buchel2, nor}.

An outline of the paper is as follows: In section \ref{screen}, we
review the calculation of the screening masses for scalar, vector and
spin-two symmetry channels in the field theory.  In section 3, we compute the leading corrections to specific screening masses. In section 4, we compare our results with QCD. Two appendices contain certain  details about the calculations used in the paper.

\section{Screening masses} \label{screen}

We will be interested in the screening masses arising due to exchange of the graviton. The graviton fluctuations can be split into scalar, vector and spin-2 symmetry
channels \cite{kovtunstarinets}. For convenience, the screening masses in the ${\mathcal N}=4$ plasma arising from 2-derivative gravity is reproduced \cite{Bak1,Brower} in table 1.   Here $\mathcal{P}$, $\mathcal{C}$
and $\mathcal{T}$ represent parity, charge conjugation and time reversal and in
the Hilbert space interpretation respectively. $\mathcal{CT}$ corresponds to the
Euclidean time reversal.

\begin{table}[hb]
\centering
\begin{tabular}{|c|c|c|c|}
\hline
SUGRA modes & $J^{PCT}$ & $m_0$ & SYM operator \\
\hline
$g_{00}$ & $0^{+++}$ &  $2.3361$  &  $T_{00}$ \\
$g_{ij}$ & $2^{+++}$ &  $3.4041$  &  $T_{ij}$ \\
$a$      & $0^{-+-}$ &  $3.4041$  &  $tr(E.B)$ \\
$\phi$   & $0^{+++}$ &  $3.4041$  &  $\mathcal{L}$ \\
$g_{i0}$ & $1^{++-}$ &  $4.3217$  &  $T_{i0}$ \\
$B_{ij}$ & $1^{+--}$ &  $5.1085$  &  $\mathcal{O}_{ij}$ \\
$C_{ij}$ & $1^{--+}$ &  $5.1085$  &  $\mathcal{O}_{30}$ \\
$B_{i0}$ & $1^{--+}$ &  $6.6537$  &  $\mathcal{O}_{i0}$ \\
$C_{i0}$ & $1^{+--}$ &  $6.6537$  &  $\mathcal{O}_{3j}$ \\
$G_a^a$  & $0^{+++}$ &  $7.4116$  &  $tr(F^4)$ \\
\hline
\end{tabular}
\caption{Spectrum \cite{Brower,Bak1}. We are interested in $g_{00}, g_{0i}$ and $g_{ij}$ in this paper.}
\label{spect}
\end{table}

Our goal is to calculate various screening masses in the deconfined
phase. Algebraically it
works out to be equivalent to working with various components of
the graviton in the soliton background as explained in \cite{Bak1}. This will also facilitate a
direct comparison of numerics with \cite{Constable} and \cite{Brower}.
The reason for this equivalence is following: we consider modes of the form $e^{i E t-i k^i z_i}$,
where $(E,k^i)$ and $(t,z^i)$ are three dimensional vectors. We get the black hole
solution by a double analytic continuation of the coordinates $\tau \to
it$ and $t \to i\tau$, where $\tau$ is the compactified spatial
dimension. So the graviton modes in the
soliton background, that are of the form $e^{i E t}$, will correspond to
exponentially decaying modes in the black hole background. With $k^i=0$, $E$ will correspond to the screening mass.
Now we briefly discuss the properties of AdS
soliton and calculate various screening masses following the discussion
in \cite{Constable}.


The 5-dimensional AdS soliton metric is written as
\begin{equation}
ds^2=\frac{r^2}{L^2}(f(r)d\tau^2+\eta_{\mu\nu}dx^{\mu}dx^{\nu}) + \frac{L^2}{r^2 f(r)}dr^2\,,
\label{soliton}
\end{equation}
\begin{equation}
\textrm{with}\quad f(r)=\left( 1-\frac{R_0^4}{r^4}\right)\,. \notag
\end{equation}
Here $\eta_{\mu\nu}dx^{\mu}dx^{\nu}=-dt^2+dx^2+dy^2$ and
$\eta_{\mu\nu}$ is the 3-dimensional Minkowski metric. Here we choose
the notation in which indices $a$, $b$ etc. will be used for the
5-dimensional spacetime and $\mu$, $\nu$ etc. for 3-dimensional
Minkowski spacetime. This geometry is constructed by double analytic
continuation of a planar AdS black hole and the conical singularity is
removed by making the $\tau$ coordinate periodic with periodicity
$
\beta_{soliton}=\frac{\pi L^2}{R_0}$.
By double analytic continuation when we get the black hole solution, the
periodicity $\beta_{soliton}$ can be related to the temperature of the
black hole
$T= 1/\beta_{soliton}$\,.

Now, to determine the screening masses corresponding to a graviton exchange,
we solve for the  linearized equation of  motion for the  graviton on
the background \eqref{soliton}. We write the perturbed metric as
\begin{equation}
g_{ab}=\bar{g}_{ab}+ \epsilon h_{ab}\,,
\label{eqn2}
\end{equation}
where $\bar{g}_{ab}$ denotes the background metric \eqref{soliton} and
$\epsilon h_{ab}$ is the perturbation with $\epsilon \ll 1$. The
background metric is a solution of the five dimensional
Einstein-Hilbert action with a negative cosmological constant
\begin{equation}
I=\frac{1}{2 \ell^3_p}\int d^5x \sqrt{-g}\left[ R +\frac{12}{L^2}\right]\,,
\label{eqn3}
\end{equation}
where $R$ is Ricci scalar, $\ell_p$ is the Plank length and $L$ is AdS
radius. We insert \eqref{eqn2} in this 5-dimensional effective action
and collect the terms second order in perturbations; {\it  i.e.},
$\mathcal{O}(\epsilon^2)$. Using this second order lagrangian we 
find the equations of motion for the perturbations. Now we make the ansatz
that the graviton has a solution of the form $h_{ab}=H_{ab}(r)e^{ik.z}$.
Here $H_{ab}(r)$ has only radial dependence and vectors $z^\mu$ and $k^\mu$ are
in 3-dimensional spacetime with $k^2=-M_0^2$; {\it  i.e.}, $k$ is the
3-dimensional momentum vector. We choose
the rest frame: $k_\mu =
(-M_0,0,0)$. Now we can solve the equations of motion by making proper
gauge choices that will also ensure that we are classifying the modes
into scalar, vector or spin-2 excitations.

\subsection{Scalar excitations}
The scalar excitations will be related to the diagonal polarization of
the graviton and also gives the mass gap in the dual field theory. We begin with
the ansatz that the solution is of the following form:
\begin{align}
H_{\tau \tau}(r)\quad &=\quad -\frac{r^2}{L^2}f(r)H_0(r)\,, \notag \\
H_{\mu \nu}(r)\quad &=\quad \frac{r^2}{L^2} \left( \eta_{\mu \nu} a_0(r)
 + \frac{k_{\mu}k_{\nu}}{M_0^2} (b_0(r)+a_0(r)) \right)\,, \notag \\
H_{rr}(r) \quad &=\quad \frac{L^2}{r^2}f^{-1}(r) c_0(r)\,, \notag \\
H_{r\mu}(r)\quad &=\quad ik_{\mu} d_0(r)\,.
\label{eqn8}
\end{align}
Here $\eta_{\mu \nu}$ is the metric on the Minkowski section of the
\eqref{soliton}. Choosing $k^{\mu}=(M_0,0)$, we get the following
simplified form of the perturbations
\begin{align}
h_{\tau \tau}\quad =&\quad -\frac{r^2}{L^2}f(r)H_0(r) e^{-iM_0t}\,,  &h_{tt}
 \quad =& \quad \frac{r^2}{L^2} b_0(r)e^{-iM_0t}\,, \notag \\
h_{ii} \quad =&\quad \frac{r^2}{L^2} a_0(r)e^{-iM_0t}\,,   &h_{rr}
 \quad =&\quad \frac{L^2}{r^2}f^{-1}(r) c_0(r)e^{-iM_0t}\,, \notag \\
h_{rt}\quad =&\quad -iM_0 d(r) e^{-iM_0t}\,,
\label{eqn8x1}
\end{align}
where now $i$ represents the spatial coordinates of the
Minkowski spacetime. As explained in \cite{Constable}, to
solve the equations of motion consistently, we also require
\begin{align}
a_0(r) \quad =& \quad \frac{H_0(r)}{2}\,, \notag \\
c_0(r) \quad =& \quad \frac{-2 R_0^{4}}{3r^{4}-R_0^{4}}H_0(r) \notag \quad \textrm{ and}\\
d_0(r) \quad =& \quad \frac{r^2}{2L^3 M_0^2}\frac{d b_0(r)}{d r}-
\frac{ r^2R_0^{4}}{(3r^{4}-R_0^{4})L^3 M_0^2}\frac{d H_0(r)}{d r} \notag \\
& \quad -\frac{12 r^{4}R_0^{4}}{(3r^{4}-R_0^{4})^2 L^2 M_0^2}H_0(r)\,.
\label{eqn9}
\end{align}

These conditions are obtained by adopting the following strategy to solve
the equations of motion:

$\bullet$ First, we make a choice of the gauge by defining $a_0 =
H_0(r)/2$.

$\bullet$ By  looking at the equations of motion of $h_{tr}$,
we see that it is satisfied if we choose $c_0(r) = -2
R_0^{4}H_0(r)/(3r^{4}-R_0^{4})$.

$\bullet$ Using these $a_0(r)$ and $c_0(r)$ we
simplify the other equations of motion. If the equation for $h_{xx}$ is subtracted from equation for $h_{\tau
\tau}$, we find the expression \eqref{eqn9} for $d_0(r)$.

$\bullet$ Now by simplifying the rest of the equations of motions, we get the
following differential equation
\begin{equation}
r (r^4-R_0^4) \frac{d^2H_0(r)}{d^2r}+(5r^4-R_0^4)\frac{d H_0(r)}{d r}+
\frac{r (64 r^2 R_0^8 + L^4 M_0^2 (-3 r^4 + R_0^4)^2)}{ (-3 r^4 + R_0^4)^2}H_0(r)=0.
\label{eqn10}
\end{equation}
Here, we point out that to solve the equations of motions we do not need
to fix $b_0(r)$. In the beginning
we could have fixed $b_0(r)$  and then the rest of the conditions
would have followed accordingly.  

The differential equation \eqref{eqn10} is solved for the value of
$M_0$ such that the solution satisfies certain boundary conditions. The
required boundary conditions are the following: first, we impose the
condition that $H_0(r)$ is finite at the horizon. This condition can be
implemented by simply solving \eqref{eqn10} near the horizon. We find that close to $r=R_0,$ the solution should
behave like
\begin{equation}
H_0(r)=1 - \frac{(16 R_0^2 +L^4 M_0^2)}{4 R_0^3}  (r - R_0).
\label{eqn11}
\end{equation}
The second boundary condition is that $H_0(r)$ should fall off
asymptotically as $1/r^4$. This condition comes from the fact that
metric fluctuations should fall off precisely with the rate to yield a
non-vanishing stress-energy tensor $\langle T_{ab}\rangle$ in the dual
field theory \cite{Myers2}. So asymptotically we expect that
\begin{equation}
H_0(r)= \frac{L^4 C_1}{r^4}+\frac{L^6 C_2}{r^6}+\frac{L^8 C_3}{r^8}\cdots \,,
\label{eqn12}
\end{equation}
where $C_1$, $C_2$, $C_3$ etc. are dimensionless constants.

To solve equation \eqref{eqn10} numerically we express
\eqref{eqn10} and \eqref{eqn11} in dimensionless variables $u$ and
$\mathcal{M}_0$, where $r=u R_0$ and $\mathcal{M}_0=L^2 M_0/R_0$. Now
we solve the differential equations iteratively and find the numerical
value of $\mathcal{M}_0$ such that the solution satisfies both the
boundary conditions.This is the shooting method.
We
find that screening mass for the scalar symmetry channel, in units of
temperature, is
\begin{equation}
M_0 =2.336\pi T\,,
\label{massgap}
\end{equation}
which is consistent with \cite{Constable, Brower}.

\subsection{Vector excitations}
To study the vector and spin-2 excitations, we start with the following
ansatz for the graviton polarizations
\begin{equation}
H_{ab}=\varepsilon_{ab} \frac{r^2}{L^2} H(r)\,.
\label{eqn13}
\end{equation}
Here, $\varepsilon_{ab}$ is a constant polarization tensor and 
satisfies
\begin{equation}
\varepsilon_{\tau a} = \varepsilon_{r a} = 0 = \varepsilon_{a \mu} k^{\mu}
 \qquad \forall a\,,
\label{eqn14}
\end{equation}
where $k^{\mu}$ is the momentum vector in the Minkowski spacetime.
Further, to separate the vector excitations, we impose the condition
that the non-vanishing components of the polarization tensor take the
following form
\begin{equation}
\varepsilon_{\mu \tau} = \varepsilon_{\tau \mu} = v_{\mu}\,, \quad \textrm{ with }
 k.v=0 \textrm{ and } v.v=1\,.
\label{eqn15}
\end{equation}
 As we have
chosen $k^\mu=(M_{v0},0,0)$, for convenience, we can choose
$v^\mu=(0,1,0)$. We write $H(r)=H_{v0}(r)$ and
with these choices, the only non-zero perturbation is
\begin{equation}
h_{\tau x}=\frac{r^2}{L^2} H_{v0}(r) e^{-i M_{v0} t}\,.
\label{eqn16}
\end{equation}
Just to clarify the notation, we point out that in $M_{v0}$ and
$H_{v0}(r)$, the subscript `$_v$' is for vector excitation and the
subscript `$_0$' is to indicate that we are working with Einstein gravity in \eqref{eqn3}. We find that the
equation of motion for this perturbation is
\begin{equation}
r \frac{d^2H_{v0}(r)}{dr^2}+5\frac{dH_{v0}(r)}{dr}+
\frac{L^4 M_{v0}^2 r}{r^4-R_0^4} H_{v0}(r)=0\,.
\label{eqn17}
\end{equation}
Now this equation of motion is solved numerically for $M_{v0}$
such that the solution satisfies proper boundary conditions. We find that close to the horizon $r=R_0$, the
solution behaves like
\begin{equation}
H_{v0}(r)=(r - R_0)\,.
\label{eqn18}
\end{equation}
The second boundary condition is that, similar to \eqref{eqn12}, the
asymptotic solution should be
\begin{equation}
H_{v0}(r)= \frac{L^4 C_1}{r^4}+\frac{L^6 C_2}{r^6}+\frac{L^8 C_3}{r^8}\cdots\,.
\label{eqn19}
\end{equation}
We can again express the equations \eqref{eqn17} and \eqref{eqn18} in
dimensionless variables and solve the equation of motion numerically.
We find that in units of temperature, the screening mass for
vector channel is
\begin{equation}
M_{v0}=4.322 \pi T\,,
\label{eqn20}
\end{equation}
in agreement with \cite{Constable, Brower}.
\subsection{Massive spin-2 excitations}
The remaining graviton fluctuations, $g_{ij}$,  describe the massive spin-2
excitations on field theory side. We again start with the ansatz
\eqref{eqn13} (with $H(r)=H_{s0}(r)$) and the polarization tensor
satisfies the conditions \eqref{eqn14}. We impose the tracelessness condition:
\begin{equation}
\eta^{\mu\nu} \varepsilon_{\mu\nu}=0\,.
\label{eqn21}
\end{equation}
Thus, now \eqref{eqn13}
will describe two independent modes, one of which will be off-diagonal, {\it  i.e.}, $\varepsilon_{xy}=\varepsilon_{yx}=1$ and otherwise
$\varepsilon_{ab}=0$. Another mode will be the  diagonal and traceless with $\varepsilon_{xx}=-\varepsilon_{yy}=1,$ and
$\varepsilon_{ab}=0$ otherwise. For convenience, we will calculate the
screening mass for the off-diagonal polarization which does not mix with other polarizations. In this case, the
only non-zero perturbations will be
\begin{equation}
h_{xy}=\frac{r^2}{L^2} H_{s0}(r) e^{-i M_{s0} t}\,,
\label{eqn21x1}
\end{equation}
and the equation of motion for it is
\begin{equation}
r (r^4 - R_0^4)\frac{d^2H_{s0}(r)}{dr^2}+(5r^4-R_0^4)
\frac{dH_{s0}(r)}{dr}+r L^4 M_{s0}^2  H_{s0}(r)=0\,.
\label{eqn22x1}
\end{equation}
We find that close to the
horizon $r=R_0$, the solution is given by
\begin{equation}
H_{s0}(r)=1 - \frac{L^4 M_{s0}^2 }{4 R_0^3} (r - R_0)\,.
\label{eqn22x2}
\end{equation}
The other boundary condition is that the solution has asymptotic behavior
similar to \eqref{eqn12}. Now we solve the equation of motion
\eqref{eqn22x1} numerically for $M_{s0}$ and get \cite{Constable, Brower}
\begin{equation}
M_{s0}=3.404 \pi T\,.
\label{eqn22x3}
\end{equation}


\section{Quantum corrections to screening} \label{correct}
As we have explained, we are interested in finding out the screening mass in the Yang-Mills plasma corresponding to a graviton exchange in the Polyakov loop correlator on the gravity side. This entails expanding the IIB low energy effective action to quadratic order in fluctuations. In the 10-dimensional action only the metric and the 5-form flux are turned on. This simplifies the problem enormously since we do not have to worry about terms such as $R H^2$ where $H$ is a combination of the RR and NSNS 3-form field strengths. One problem that we could face is that there could be mixing between the metric fluctuation and the scalar or the gauge field. Since we do not know these terms at higher derivative order completely, our results could be incomplete. Now we will argue that such mixing will not arise. Firstly, we assume that since we are in a perturbative approximation, there is no level crossing,{\it  i.e.}, the lightest mode remains the lightest mode and so on. When we Fourier expand, each mode comes with a factor of $e^{-i M_i t}$ where $M_i$ is the eigenvalue corresponding to that mode. Thus integration over time is only going to allow for mixing at quadratic order between degenerate modes. Looking at table 1, it is thus clear that the only problematic terms could arise due to the mixing of the metric and the dilaton-axion at quadratic order which would effect the result for the spin-2 exchange. Firstly note that the dilaton is constant to leading order. It gets sourced by the leading higher derivative correction at $O(\gamma)$. Thus any fluctuation should also be at least $O(\gamma)$. Thus plugging this into the action would lead to contributions at least at $O(\gamma^2)$. Since there is an SL(2,Z) symmetry, we would also conclude that the axion would contribute at $O(\gamma^2)$. We can do somewhat better than this.  Since there is an SL(2,Z) symmetry in IIB string theory, a term of the form $\partial^m h \partial^n \phi$ (where $h$ and $\phi$ denote metric and dilaton fluctuations around the background) would necessitate the existence of $\partial^m h \partial^n a$ where $a$ is an axion-fluctuation. The 10-dimensional origin of the latter term could be of the form $C_{abcd}C^{aec}_{\ \ \ f} C^{bd}_{\ \ e g} \nabla^f \nabla^g C^{(0)}$. However such terms cannot arise in perturbation theory due to parity conservation. As the problematic terms alluded to above will not arise and we need to simply focus on the curvature corrections. At this point, we must point out that although we are guided by type IIB string theory, the phenomenological program in \cite{Buchel1} is more general and we expect corrections starting at four derivative order with unknown mixings between various fields. Nevertheless, if we work in a perturbative approximation, our comments above will also apply to the general case.

Now 
the most well known curvature correction in IIB is the $W^4$ term with a specific contraction between 4 Weyl tensors dictated by supersymmetry. In general, we may expect that due to addition of fundamental matter, the corrections would begin at $W^2$ order and in a non-supersymmetric plasma there will also be $W^3$ terms and a more general set of $W^4$ terms. We will not consider terms involving covariant derivatives of the Weyl tensor in the non-supersymmetric case--this is an assumption for simplicity. The various tensor contractions of the Weyl tensors are shown in table 2. In the next section we will show in detail the analysis with $W^4$ which is relevant for the screening masses in the $\mathcal{N}=4$ plasma.

\begin{table}[hb]
\centering
\begin{tabular}{ | c | c | c | c |  }
  \hline
  &$W^4_i$ terms & $W^3$ terms & $W^2$ terms \\
  \hline
  $1.$ &$W^{wvrs}W_{wv}$$^{tu}W_{rmt}$$^{n}W_{snu}$$^{m}$ & $W^{rstu}W_{rt}$$^{vw}$$W_{svuw}$ & $W_{rstu}$$W^{rstu}$  \\
  \hline
  $2.$&$W^{wvrs}W_{wvr}$$^{t}W_{s}$$^{umn}W_{tmun}$ &  & \\
  \hline
  $3.$&$W^{wvrs}W_{wr}$$^{tu}W_{vt}$$^{mn}W_{smun}$ &  & \\
  \hline
  $4.$&$W^{wvrs}W_{wtr}$$^{u}W_{v}$$^{mtn}W_{snum}$ &  & \\
  \hline
  $5.$&$W^{wvrs}W_{wtr}$$^{u}W_{vms}$$^{n}W^{tm}$$_{un}$ &  & \\
  \hline
\end{tabular}
\caption{Independent contractions of Weyl tensors in five dimensions}
\label{table1}
\end{table}

Since we wish to consider general curvature corrections as in table 2, it is instructive to review the supersymmetric $W^4$ case first in detail to set out the procedure for the remaining cases.
We begin with the 5-dimensional effective action in  Einstein frame including the eight derivative correction term \footnote{These terms originate from the dimensional reduction of the 10-dimensional action consisting of the metric and the five-form flux \cite{paulos,quantum, Buchel3}.} $W_s^4$
\begin{equation}
I=\frac{1}{2 \ell_p^3}\int d^5x \sqrt{-g}\left[ R +\frac{12}{L^2}+\gamma L^6 W_{s}^4 \right]\,,
\label{eqn22}
\end{equation}
where dimensionless coupling constant  \cite{quantum} $\gamma=\zeta(3)/8\lambda^{3/2}+\sqrt{\lambda}/384 N_c^2 \ll 1$ with $\lambda$ being the 
't Hooft coupling\footnote{Note that if $\lambda=6\pi, N_c=3$ corresponding to $\alpha_s=0.5$, then $\gamma\approx 0.003$.}   and $W_{s}^4$ is
given by
\begin{equation}
W_{s}^4=W^{hmnk}W_{pmnq}W_{h}^{\;\;rsp}W^{q}_{\;\;rsk}+\frac{1}{2}W^{hkmn}W_{pqmn}W_{h}^{\;\;rsp}W^{q}_{\;\;rsk}\,.
\label{eqn23}
\end{equation}
This form of the corrections appears in the supergravity action and 
is equal to $(1/2 W^4_1+ W^4_4)$, where $W^4_i$ is the $i^{th}$-contraction of four Weyl tensors in Table \ref{table1}. 

We write the soliton solution as
\begin{equation}
ds^2=\frac{r^2}{L^2}(f(r)d\tau^2+\eta_{\mu\nu}dx^{\mu}dx^{\nu}) + \frac{L^2}{r^2 g(r)}dr^2\,,
\label{eqn23x1}
\end{equation}
with
\begin{align}
f(r)\quad =& \quad f_0(r)(1+\gamma f_1(r))\,, \notag \\
g(r)\quad =& \quad f_0(r)(1+\gamma g_1(r)) \notag \\
\textrm{and}\quad f_0(r)\quad =& \quad \left( 1-\frac{R_0^4}{r^4}\right).
\label{eqn23x2}
\end{align}
We plug in this solution into the action \eqref{eqn22} and find
equations of motion for $f_1(r)$ and $g_1(r)$. After solving these
coupled differential equations we find that
\begin{align}
g_1(r)\quad =& \quad \frac{C_1 r^{12} + 360 r^4 R_0^{12} - 285 R_0^{16}}{r^{12} (r^4 - R_0^4)}\,, \notag \\
f_1(r)\quad =& \quad  \frac{C_1 r^{12} +120 r^4 R_0^{12} -45 R_0^{16}}{ r^{12} (r^{4} -R_0^4)} + C_2\,,
\label{eqna23x3}
\end{align}
where $C_1$ and $C_2$ are constants.  To fix these constants we impose
following conditions:

$\bullet$ We demand that the horizon is still at $r=R_0$. So $g_1(r)$
and $f_1(r)$ should be regular at $r=R_0$, {\it  i.e.}, the numerator of the
expression for $g_1(r)$ should vanish at $r=R_0$. We get
\begin{align}
C_1 \quad =& \quad -75 R_0^4 \notag \\
\textrm{and now}\quad g_1(r)\quad =& \quad \frac{-15 R_0^4 (5 r^8 + 5 r^4 R_0^4 - 19 R_0^8)}{r^{12}}\,, \notag \\
f_1(r)\quad =& \quad  \frac{-15 R_0^4 (5 r^8 + 5 r^4 R_0^4 - 3 R_0^8)}{r^{12}}+C_2\,.
\label{eqn23x4}
\end{align}
Here we notice that by choosing the given $C_1$, the stated condition
is satisfied by both $g_1(r)$ and $f_1(r)$.

$\bullet$ To fix $C_2$ we use the following condition: to find black
hole solution of the action \eqref{eqn22} we just need to do the double
analytic continuation in the metric \eqref{eqn23x1}; i.e., $\tau \to
it$ and $t \to i\tau$. In this solution, we can extract the background
metric for the dual gauge theory by going to the asymptotic limit giving
\begin{equation}
ds^2=-f(\infty)dt^2+d\tau^2+dx^2+dy^2\,.
\label{eqn23x5}
\end{equation}
To have the speed of light to be one in the dual gauge theory, one
requires $f(\infty)=1$ and this fixes the constant $C_2=0$.

Having found the soliton solution, we remove the conical singularity by
compactifing the $\tau$ coordinate to be periodic and this periodicity
$\beta_{soliton}$ will be related to the temperature of the dual field
theory
\begin{equation}
T=\frac{1}{\beta_{soliton}}=\frac{R_0 }{L^2 \pi}(1 +15 \gamma)\,.
\label{eqn23x6}
\end{equation}

Now to calculate the free energy of the dual gauge theory, we use
standard path integral technique in which we identify the Euclidean
action ($I_{euc}$) of the bulk gravity with the ratio of the free
energy and temperature ($w/T$) of the dual field theory. To render the
action finite, following \cite{Gubser1}, we use background subtraction to compute the free energy:
\begin{equation}
w=-\frac{\pi^4 L^3 T^4}{2 \ell_p^3} ( 1+15\gamma)\,.
\label{eqn23x7}
\end{equation}
leading to the entropy density:
\begin{equation}
s=-\frac{\partial w}{\partial T}=\frac{2\pi^4 L^3 T^3}{ \ell_p^3} ( 1+15\gamma)\,.
\label{eqn23x8}
\end{equation}

\subsection{Scalar excitation}
Having studied the thermodynamic properties of this perturbed solution
we now turn to computing the correction to the mass gap. We begin with
\begin{equation}
g_{ab}=\bar{g}_{ab}+\epsilon h_{ab}\,,
\label{eqn23x10}
\end{equation}
where $\epsilon \ll 1$. We insert \eqref{eqn23x10} in the effective
action \eqref{eqn22} and find the terms of the lagrangian that are
second order in $\epsilon$ and use these terms to find equations of
motion for perturbations. Now we make the ansatz that the solution is of the form
$h_{ab}=H_{ab}e^{-ik.z}$, where $k$ and $z$ are three-dimensional
vectors in the Minkowski section of metric \eqref{eqn23x1}. Here
$z^{\mu}=(t,x,y)$ and $k$ is momentum vector with $k^2=-M^2$. As we
want to find the correction to the mass gap, we expect that up to first
order $M=M_0+\gamma M_1$, where $M_0$ is screening mass for the scalar
excitation given by \eqref{massgap} and $M_1$ is the first order
correction. We further make the ansatz that solution is of the form:
\begin{align}
H_{\tau \tau}(r)\quad &=\quad -\frac{r^2}{L^2}f(r)H(r)\,, \notag \\
H_{\mu \nu}(r)\quad &=\quad \frac{r^2}{L^2} \left( \eta_{\mu \nu} a(r) + \frac{k_{\mu}k_{\nu}}{M^2} (b(r)+a(r)) \right)\,, \notag \\
H_{rr}(r) \quad &=\quad \frac{L^2}{r^2}f^{-1}(r) c(r)\,, \notag \\
H_{r\mu}(r)\quad &=\quad ik_{\mu} d(r)\,.
\label{eqn23x11}
\end{align}
 For convenience, we choose the frame where
$k^{\mu}=(M,0,0)$ so that the solution is of the form
$h_{ab}=H_{ab}e^{-iMt}$. These expressions are similar to \eqref{eqn8}
but now all the terms have first order corrections in $\gamma$. So we
write
\begin{align}
H(r) \quad =& \quad H_0+\gamma H_1\,,  &a(r) \quad =& \quad a_0(r)+\gamma a_1(r)\,, \notag \\
b(r) \quad =& \quad b_0(r)+\gamma b_1(r)\,, &c(r) \quad =& \quad c_0(r)+\gamma c_1(r)\,, \notag \\
d(r) \quad =& \quad d_0(r)+\gamma d_1(r)\,, & M \quad =&\quad M_0+\gamma M_1\,,
\label{eqn23x12}
\end{align}
where $H_0$, $a_0$, $b_0$, $c_0$ and $d_0$ are zeroth order solutions.
The perturbations  are of the following form
\begin{align}
h_{\tau \tau}\quad =&\quad -\frac{r^2}{L^2}f(r)H(r) e^{-iMt}\,,
& h_{tt} \quad =& \quad \frac{r^2}{L^2} b(r)e^{-iMt}\,, \notag \\
h_{\mu \mu} \quad =&\quad \frac{r^2}{L^2} a(r)e^{-iMt}\,,
&h_{rr} \quad =&\quad \frac{L^2}{r^2}f^{-1}(r) c(r)e^{-iMt} \,, \notag \\
h_{rt}\quad =&\quad -iM d(r) e^{-i M t}\,.
\label{eqn23x13}
\end{align}

The zeroth order terms of the equations of motion ($\mathcal{O}(\gamma^0)$) give
the differential equation \eqref{eqn10}. The first order terms
of equations of motion will be functions of $H_0(r)$, $H_1(r)$,
$a_1(r)$, $b_1(r)$, $c_1(r)$, $d_1(r)$, $M_0$, $M_1$ and their
derivatives. Now we explain how we solve the first order equations of
motion for perturbations:

$ \bullet$ First, we eliminate the higher derivatives of $H_0(r)$ in the
first order equations of motion ($\mathcal{O}(\gamma)$) by using the
leading order equation of motion \eqref{eqn10} and its derivatives.

$ \bullet $ Similar to the discussion in section 2, we fix the gauge by
choosing $a_1(r)=H_1(r)/2$.

$ \bullet $ Now, we find the expression for $c_1(r)$ by demanding that
the equation of motion for $h_{tr}$ is satisfied. We have given the
expression for $c_1(r)$ in the appendix \ref{appA}. It can be seen that
the functional dependence of $c_1(r)$ on $H_1(r)$ is similar to the
dependence of $c_0(r)$ on $H_0(r)$ in \eqref{eqn9}. There are extra
terms in expression of $c_1(r)$ that depend on $H_0(r)$ and these terms
come from the correction $\gamma W^4_s$.

$ \bullet$ After using this expression for $c_1(r)$, we find that the
equation for $h_{tt}$ is also satisfied. Now we use the
equation for the perturbation $h_{rr}$ and find the functional
form of $d_1(r)$. We give the explicit expressions in appendix
\ref{appA}.

$\bullet$ After using this form of $d_1(r)$, the equation for
$h_{rr}$ is satisfied and from rest of the equations of motion we get
the differential equation for $H_1(r)$. Similar to the leading order solution, we fix the
gauge by fixing either $a_1(r)$ or $b_1(r)$ and rest of the consistency
conditions follow accordingly. We also notice that the differential
equation for $H_1(r)$ (equation \eqref{eqn51} in the appendix
\ref{appA}) is similar to \eqref{eqn10}, but with the leading order
solution $H_0(r)$ behaving as source.

$\bullet$ Equations \eqref{eqn10} and the equation for $H_1(r)$
are coupled differential equations. Now, knowing the value of $M_0$
from \eqref{massgap}, we can solve for $M_1$. The boundary conditions for
$H_0(r)$ is given by \eqref{eqn11} and \eqref{eqn12}.
Similar to the boundary condition for $H_0(r)$, we solve the equation
of motion for $H_1(r)$ (equation \eqref{eqn51}) near the horizon $r=R_0$. $H_1(r)$ near the horizon is given by \eqref{eqn23x16}. Similar to $H_0(r)$ in \eqref{eqn12}, the second boundary condition for the solution is that asymptotically $H_1(r)$ falls off as $1/r^4$.

Finally, using the shooting method we find
\begin{align}
M=\pi T (2.336-139.514 \gamma)\,.
\label{eqn23x17}
\end{align}

A similar analysis can be done for all the perturbations given in Table
\ref{table1}. We have show the results for these
terms in the Tables \ref{table4}, \ref{table5}, \ref{table2} and
\ref{table3}.

\subsection{Vector excitation}

In this section, we calculate the correction to the screening mass for
the vector channel.
We start with a perturbed metric of the form \eqref{eqn23x10}. We
insert this metric into the effective action \eqref{eqn22} and find the
equations of motion for perturbations by singling out the terms
$\mathcal{O}(\epsilon^2)$. Now we make the ansatz that
$h_{ab}=H_{ab}e^{-ik.z}$, where $k$ and $z$ are three-dimensional
vectors in the Minkowski metric \eqref{eqn23x1} with $k^2=-M_v^2$. We
further expect that $M_v=M_{v0}+\gamma M_{v1}$, where $M_{v1}$ is first
order correction to the screening mass. We further make the ansatz that
the solution is of the following form
\begin{equation}
H_{ab}=\varepsilon_{ab} \frac{r^2}{L^2} H_v(r)\,.
\label{eqn24x1}
\end{equation}
Here $\varepsilon_{ab}$ is a constant polarization tensor and satisfies \eqref{eqn14} and \eqref{eqn15}.
The only non-zero perturbation is
\begin{equation}
h_{\tau x}=\frac{r^2}{L^2} H_{v}(r) e^{-i M_v t}\,.
\label{eqn24x3}
\end{equation}
For this perturbation, we assume that $H_{v}(r)=H_{v0}(r)+\gamma
H_{v1}(r)$, where $H_{v0}$ is the solution of differential equation
\eqref{eqn17} and $H_{v1}$ is the first order correction.

To find the equation of motion for $H_{v1}(r)$, we plug in
\eqref{eqn24x3} in the equation of motion for the perturbation $h_{\tau
x}$. After inserting $M_v=M_{v0}+\gamma M_{v1}$, we expand this
equation of motion in powers of $\gamma$. We see that the zeroth order
terms in $\gamma$ will give us the equation of motion for $H_{v0}(r)$,
{\it i.e.}, equation \eqref{eqn17}. We can use this zeroth order equation of
motion to eliminate derivatives of $H_{v0}(r)$ from terms first order
in $\gamma$. Finally, we get equation of motion for $H_{v1}(r)$ that is
given in the appendix \ref{appA} (equation \eqref{eqn51x1}).

The equation of motion for $H_{v1}(r)$ and $H_{v0}(r)$ are coupled
differential equations. Now we can find $M_{v1}$ by using the value of
$M_{v0}$ (equation \eqref{eqn20}) and solving this set of coupled
differential equations such that $H_{v1}(r)$ satisfies proper boundary
conditions. Using \eqref{eqn18} we can find the solution for $H_{v1}(r)$ near the horizon and it is given by \eqref{eqn24x4}.

The second boundary condition for $H_{v1}(r)$ is that it falls off asymptotically
similar to $H_{v0}(r)$, {\it i.e.}, a function similar to \eqref{eqn19}. Now
we can solve the coupled differential equations \eqref{eqn17} and
\eqref{eqn51x1} for $M_{v1}$ by converting it into dimensionless
parameters (see discussion before \eqref{eqn23x17}). Finally, we find
that the screening mass for the vector channel is
\begin{equation}
M_v=\pi T (4.322- 398.354 \gamma)\,.
\label{eqn24x5}
\end{equation}

\subsection{Massive spin-2 excitations}
We start with $h_{ab}=H_{ab}e^{-ik.z}$. We also assume that $M_s=M_{s0}+\gamma M_{s1}$, where
$M_{s1}$ is first order correction to the screening mass. Here $M_{s0}$
is given by \eqref{eqn23x1}. Further, we make the ansatz that the
solution is of the following form
\begin{equation}
H_{ab}=\varepsilon_{ab} \frac{r^2}{L^2} H_s(r)\,.
\label{eqn24x6}
\end{equation}
Here, $\varepsilon_{ab}$ is a constant polarization tensor satisfying \eqref{eqn14} and \eqref{eqn21}.
 Now as before we focus on the
off-diagonal mode and set $\varepsilon_{xy}=\varepsilon_{yx}=1$ and
$\varepsilon_{ab}=0$ otherwise. So in this case, only non-zero perturbation
will be given by
\begin{equation}
h_{xy}=\frac{r^2}{L^2} H_{s}(r) e^{-i M_{s} t}\,.
\label{eqn24x8}
\end{equation}
We further assume that $H_{s}(r)=H_{s0}(r)+H_{s1}(r)$ and insert
\eqref{eqn24x8} into the equation of motion for $h_{xy}$. Now we expand
this differential equation in $\gamma$ and find that zeroth order terms
give us equation of motion for $H_{s0}(r)$, that is given by
\eqref{eqn22x1}. The first order terms will give the equation of motion
for $H_{s1}(r)$ and the full expression is given in the appendix
\ref{appA} (equation \eqref{eqn51x2}).

Similar to other screening masses, knowing the value of $M_{s0}$, we
solve the coupled differential equations \eqref{eqn51x2} and
\eqref{eqn22x1} for $M_{s1}$. The boundary condition for $H_{s0}$ is
that near the horizon, $H_{s0}$ is given by \eqref{eqn22x2} and it 
falls off as \eqref{eqn12}. $H_{s1}(r)$  also falls of as $1/r^4$ asymptotically. Also, we can use \eqref{eqn22x2} and find its solution near the horizon, which is give by \eqref{eqn24x9}.

Given these boundary conditions and knowing the value of
$M_{s0}$, we can solve the coupled differential equations iteratively
for the correction $M_{s1}$. Finally we find that the screening mass for the spin-2 channel is given by
\begin{align}
M_{s}=\pi T (3.404- 167.619 \gamma)\,.
\label{eqn24x10}
\end{align}

\section{Comparison with QCD} \label{qcd}

In this section, we tabulate the results obtained using the procedures outlined in section 3 and appendix B for the screening masses and $\eta/s$. Using these results we would like to make a phenomenological study of certain properties of the QCD plasma. By adding higher derivative corrections we have introduced extra parameters in the theory. In keeping with the general strategy in  \cite{Buchel1}, we will use input from lattice QCD for the energy density, the mass gap (and $M_s$) to fix these parameters. Using the fixed parameters, we will ``predict" the value for $\eta/s$ and $M_v$. Of course our results should be taken with a grain of salt since the corrections we have added are not the most general (since we have ignored covariant derivatives acting on the curvature tensor at six and eight derivative order). However, interestingly we will find that the predictions are reasonable giving hope that enlarging the space of couplings may eventually lead to sensible phenomenology. From the
latest lattice calculations, the ratio of energy density with its free field limit 
for 4-dimensional lattice QCD at temperature $T \approx 2T_c$ is $(\varepsilon/\varepsilon_0)_{lattice}=0.85-0.90$ \cite{petreczky, Cheng1}.  The values of screening masses for 2-flavour $N_f=2$ QCD at $2T_c$ are : $(M)_{lattice}/\pi T =1.68-1.91$; vector and spin-2 symmetry channels are $(M_v)_{lattice}/\pi T =2.76-3.02$ and $(M_s)_{lattice}/\pi T =2.48-2.64$ \cite{Bak1, petreczky, Maezawa}.

For comparison with lattice  results, let us consider the following action with up to eight-derivative corrections
\begin{equation}
I\,=\,\frac{1}{2 \ell_p^3}\int d^5x \sqrt{-g}\Bigg[ R +\frac{12}{L^2} + \alpha L^2 W_{abcd}W^{abcd} + \beta L^4 W^{abcd}W_{ac}^{\;\; ef}W_{bedf}+\gamma_i L^6 W_i^4 \Bigg]\,,
\label{eqn45x1}
\end{equation}
where $\alpha$, $\beta$ and $\gamma_i$ are dimensionless coupling constants and they are systematically suppressed by the powers of $\lp^2/L^2$. In $\gamma_i$, $i$ runs from 1 to 5 to give the general 8-derivative terms corresponding to table 2. This lagrangian will give a description of supersymmetric plasma if $\beta=0$ and $\gamma_i W_i^4=\gamma W^4_s$, where $W^4_s$ is given by \eqref{eqn23}. For a non-supersymmetric plasma, $\beta \neq 0$ and we also do not know the precise form of the $\gamma_i W_i^4$ term in this case.

\begin{table}[hb]
\centering
\begin{tabular}{ | c | c | c | c | c | }
  \hline
  Term  & $M_v/\pi T$  & $M_s/\pi T$ \\
  \hline
  $W^2$ & $4.322+12.965 \alpha - 114.522 \alpha^2$ & $3.404 -6.414 \alpha + 128.620  \alpha^2$\\
  \hline
  $W^3$ & $4.322+45.738 \beta$ & $3.404-25.839\beta$\\
  \hline
  $W^4_1$ & $4.322+603.164\gamma_1$ & $3.404+ 279.749 \gamma_1$\\
  \hline
  $W^4_2$  & $4.322+238.119\gamma_2 $ & $3.404-532.634 \gamma_2 $\\
  \hline
  $W^4_3$ & $4.322-285.403\gamma_3$ & $3.404-122.103 \gamma_3 $\\
  \hline
  $W^4_4$ & $4.322-695.936\gamma_4$ & $3.404-307.493 \gamma_4 $ \\
  \hline
  $W^4_5$ & $4.322-216.641\gamma_5 $ & $3.404-673.803 \gamma_5 $\\
  \hline
\end{tabular}
\caption{Corrections to $M_v, M_s$}
\label{table4}
\end{table}

\begin{table}
\centering
\begin{tabular}{ | c | c | c | c | c | }
  \hline
  Term & $4\pi  \eta/s$ & Mass gap$/\pi T$\\
   & &  \\
  \hline
  $W^2$ & $(1-8 \alpha+112 \alpha^2)$ & $2.336 +  17.066 \alpha - 97.704 \alpha^2$\\
  \hline
  $W^3$ & $(1-96 \beta)$ & $2.336- 13.450\beta$\\
  \hline
  $W^4_1$ & $(1-416\gamma_1)$ & $2.336+ 266.317 \gamma_1$\\
  \hline
  $W^4_2$  & $(1+832\gamma_2)$ & $2.336- 532.634 \gamma_2 $\\
  \hline
  $W^4_3$ & $(1+120\gamma_3)$ & $2.336- 139.515 \gamma_3 $\\
  \hline
  $W^4_4$ & $(1+328\gamma_4)$ & $2.336- 272.673 \gamma_4 $ \\
  \hline
  $W^4_5$ & $(1+688\gamma_5)$ & $2.336- 691.218 \gamma_5 $\\
  \hline
\end{tabular}
\caption{Corrections to $\eta/s$ and mass gap}
\label{table5}
\end{table}

Implicitly in displaying these results we have assumed 
$O(\alpha^2)\sim O(\beta)$, 
$O(\alpha^3)\ll O(\beta)$, $O(\alpha^3)\ll O(\gamma_i)$ and $O(\beta^2)\ll O(\gamma_i)$. These are respected by the numerical solutions we have obtained in the discussion. Futhermore, we have included the $\alpha^2$ terms to check the consistency of our numerics.

Let us start by assuming $\gamma_i W_i^4=\gamma W_s^4$; we will soon comment on the general form.
Now, from table \ref{table3}, we can see that energy density for this lagrangian is given by
\begin{align}
\varepsilon = \frac{3 \pi^4 L^3 T^4}{2\ell_p^3}(1 + 18 \alpha + 24 \alpha^2+ 6 \beta + 15 \gamma)\,.
\label{eqn45x1x1}
\end{align}
We would like to take its ratio with the energy density of a non-supersymmetric conformal plasma in free field limit. To find energy density for free fields, we begin with the comparison of conformal anomalies of a four dimensional CFT with the one calculated using
holographic techniques \cite{Buchel1,Buchel2}:
\begin{align}
a=\pi^2 \frac{L^3}{\ell_p^3} \quad \textrm{and} \quad c= \pi^2 \frac{L^3}{\ell_p^3}(1+8\alpha)\,.
\label{eqn45x1x2}
\end{align}
Here $a$ and $c$ are central charges of CFT. Further, if we consider a free 
massless 
field theory with $N_1$ vectors, $N_0$ scalars and $N_{1/2}$ chiral fermions, we find that \cite{Hofman, Birrell}
\begin{align}
a \quad =& \quad \frac{124 N_1 + 11 N_{1/2}+2 N_0}{720}\,, \label{eqn45x1x3} \\
c \quad =& \quad \frac{12 N_1 + 3 N_{1/2} + N_0 }{120}\,, \\
t_4 \quad =& \quad \frac{15(N_0 + 2 N_1 - 2 N_{1/2})}{2 (N_0 + 12 N_1 + 3 N_{1/2})}\,.
\end{align}
Here $t_4$ is a constant that characterize the three-point function in CFT. This constant is found to be related to $\beta$ by $t_4 = 4320 \beta$.
The original calculations for this term was done in \cite{Hofman} in absence of the quadratic terms and later it was argued in
\cite{Buchel2} that even in the presence of $W^4_s$ terms, the contribution to $t_4$ is of the form $\mathcal{O}(\beta \gamma,\;
\gamma^2)$. So the expression for $t_4$ is correct so long we are considering only first order corrections from $W^3$ and $W^4_s$ terms. 
We are going to normalize by the energy density for a collection of free bosons and fermions which is given by
\begin{align}
\varepsilon_0 = \frac{ \pi^2 T^4}{30}\left( N_b+\frac{7}{8}N_f \right)\,,
\label{eqn45x1x5}
\end{align}
where $N_b$ and $N_f$ are the  number of bosonic and fermionic degrees of freedom. For a non-supersymmetric theory $N_b\neq N_f$ where $N_b=N_0+2N_1$ and $N_f=2N_{1/2}$. Now for the comparison of energy density \eqref{eqn45x1x1} with its free field limit, we express quantities in \eqref{eqn45x1x5} in terms of bulk gravity parameters using \eqref{eqn45x1x2} - \eqref{eqn45x1x5}: 
\begin{equation}
\varepsilon_0 = \frac{ \pi^2 T^4}{30}\left( N_0 +2 N_1 + \frac{7}{4}N_{1/2}\right) =\frac{2 \pi^4 L^3 T^4}{\ell_p^3} \left( 1 + 16 \alpha + 192 \beta \right)\,.
\label{eqn45x1x6}
\end{equation}
Now we can use \eqref{eqn45x1x1}, \eqref{eqn45x1x6} and results from
tables \ref{table4}, \ref{table5}, \ref{table3} to find that up to
first order in $\beta$, $\gamma$ and second order in $\alpha$
\begin{align}
\frac{M}{\pi T}\quad=&\quad 2.34 + 17.07 \alpha - 97.70 \alpha^2 -13.45 \beta - 139.51 \gamma\,, \label{eqn45x2} \\
\frac{M_v}{\pi T}\quad=&\quad 4.32 + 12.97 \alpha - 114.52 \alpha^2 + 45.74  \beta - 394.35 \gamma\,, \label{eqn45x2x1} \\
\frac{M_s}{\pi T}\quad=&\quad 3.40 - 6.41 \alpha + 128.62 \alpha^2 - 25.84 \beta - 167.62 \gamma\,, \label{eqn45x2x2} \\
\frac{\varepsilon}{\varepsilon_0} \quad=&\quad \frac{3}{4}(1 + 2 \alpha - 8 \alpha^2 - 186 \beta + 15 \gamma)\,, \label{eqn45x3} \\
\frac{\eta}{s}\quad=&\quad\frac{1}{4\pi}(1 - 8 \alpha + 112 \alpha^2 -96 \beta + 120 \gamma)\,. \label{eqn45x4}
\end{align}

To begin with, let us compare spectrum of type IIB theory with QCD to see what do we get. In this case, the supersymmetric plasma only has the $W_s^4$ correction and demanding $\epsilon/\epsilon_0=[0.85,0.90]$ leads\footnote{This value of $\gamma$ would correspond to $\alpha_s=0.14-0.20$ for $N_c=3$.} to $\gamma=[0.008,0.013]$. This leads to $\eta/s=[0.16,0.20]$, $M/\pi T=[0.47,1.10]$, $M_v/\pi T=[-0.94,0.81]$ and $M_s/\pi T=[1.16, 1.91]$. Thus while $\eta/s$ seems to be in the right ball park as RHIC and LHC data show \cite{heinz}, the screening masses are underestimated. In fact for the lower limit, $M_v$ is also negative which indicates the possibility of not having a mass gap in the theory. Of course there is no reason for anything sensible with the leading $W^4$ correction which is relevant for the supersymmetric $\mathcal{N}=4$ plasma. Thus we should enlarge the space of couplings. First, we could consider adding fundamental matter which corresponds to turning on the $W^2$ term. In this case, we can use the lattice free energy and mass gap results to fix $\gamma,\alpha$ to get $M_v/\pi T = [1.03, 2.40]$, $\eta/s \in [0.123, 0.154]$ and $M_s/\pi T \in [1.87, 2.46]$. This is already an improvement! However, note that certain values of $\alpha$ and $\gamma$, which produce $M_v<M$ or $M_s<M$, are not allowed as they violate our no level-crossing assumption. So let us now consider turning on $W^3$ and more general $W^4$ terms which would be relevant for a non-supersymmetric plasma to see what we get. 

In Table \ref{table4}, \ref{table5} and \ref{table3}, we have mentioned the results for individual correction terms that might arise in the
gravitational dual of  non-supersymmetric field theories. Here we observe the following pattern: consider just the general set of $W^4$ terms. If the correction to the temperature is $T=\pi L^2/R_0 (1+ c_i \gamma_i)$ then the corrected mass gap is $M/\pi T \approx (2.34-9.4 c_i \gamma_i)$. Here $\gamma_i$ is a dimensionless coupling constant and $c_i$ depends on the different $W^4_i$ terms. Note that the above relation for the mass gap is approximate which we will use to get a range for the other physical quantities. Further, we also have the energy density $\varepsilon/ \varepsilon_0=3/4(1+ c_i \gamma_i)$ and the screening mass for spin-2 channel is $M_s/\pi T \approx (3.40-9.3 c_i \gamma_i)$. For the screening mass in the vector symmetry channel, we find that  $M_v/\pi T = 4.32+ m_1 c_i \gamma_i$ where $m_1$ varies between $-4.2$ and $24.0$ for different cases. The corrections to $\eta/s$ have similar qualitative behaviour and we find that the numerical coefficients of $ c_i \gamma_i$ can vary between 8.0 and 14.9. Now using these approximate results, we can reinstate the other corrections and compute the range for the coupling constants. To do that, similar to \eqref{eqn45x2} - \eqref{eqn45x4}, we can write these physical quantities with contribution from all the $W_i$ terms weighted by their respective coupling constants $\gamma_i$. In these expressions, we can define $\Gamma= c_i \gamma_i$. We replace $\gamma \to \Gamma/15$ in \eqref{eqn45x2}, \eqref{eqn45x2x2} and \eqref{eqn45x3}. In \eqref{eqn45x2x1} and \eqref{eqn45x4}, we replace $394.35 \gamma \to m_1 \Gamma$ and $120 \gamma \to m_2 \Gamma$.
Here we have $m_1 \in [-4.2, 24.0]$ and $m_2 \in [8.0,14.9]$. With these approximations, we have reduced the five coupling constants $\gamma_i$ to only one, namely $\Gamma$. 
Now we can see that there are uncertainties in the screening mass for the vector channel because of $m_1$, so we will compare $M$, $M_s$ and $\varepsilon/\varepsilon_0$ with the lattice calculations. Then we calculate $M_v$ and $\eta/s$, and compare these with lattice results.

By using the values of $(M)_{lattice}$, $(M_s)_{lattice}$ and $(\varepsilon/\varepsilon_0)_{lattice}$ in equations \eqref{eqn45x2}, \eqref{eqn45x2x2} and \eqref{eqn45x3} we get: 
\begin{equation}
\alpha  \;\in\; [0.00423, 0.02847]\,, \quad
\beta \;\in\; [-0.00060, 0.00003]\,, \quad
\Gamma  \;\in\; [0.07300, 0.09179]\,. 
\label{eqn45x8}
\end{equation}
Using \eqref{eqn45x8} in  \eqref{eqn45x2x1} and \eqref{eqn45x4}, we find that\footnote{ Dropping the $\alpha^2$ terms leads to a $10\%$ change in the lower range or $\eta/s$ and a $2\%$ change in the upper range of $M_v$ leaving everything else the same. }
\begin{align}
\frac{\eta}{s} & \;\in \; [0.11494, 0.19049]\,, \label{eqn47} \\
\frac{M_v}{\pi T} & \;\in \; [2.14224, 4.98317]\,. \label{eqn47x1}
\end{align}

Thus $\eta/s$ is still well within\footnote{Conservatively, for a RHIC/LHC plasma, $\eta/s<0.4$.} experimental limits \cite{heinz}. Interestingly, the range for screening mass in vector symmetry channel also encompasses the lattice results $(M_v)_{lattice}/\pi T =2.76-3.02$. This gives hope that enlarging the space of couplings along the lines of \cite{Buchel1, Buchel2} may yield physically interesting results. In our numerics, we find that $O(\beta)\sim O(\alpha^2) \ll O(\gamma)$. Curiously, the dominant contribution comes from the $\gamma$-dependent  terms.


\section{Discussion} \label{discuss}

In this paper we have studied corrections to certain screening masses in the Yang-Mills plasma at strong coupling using AdS/CFT. On the gravity side the screening masses corresponded to the exchange of a graviton in the correlator of Polyakov loops. We were able to extract corrections to the mass gap which is the lightest supergravity mode. In addition we also computed corrections to the screening masses arising in the vector and spin-2 channels. We expanded the space of couplings by adding higher derivative corrections corresponding to the addition of fundamental matter and breaking of supersymmetry. Using lattice input for the mass gap and $M_s$, we were able to get sensible predictions for $\eta/s$ and $M_v$. 

Our approach may give the impression that all that one needs to do to find a realistic model is to add new couplings and adjust them to agree with observables available to us. It would be somewhat unsatisfying if that was all that could be said about this program. We would like to make the following interesting observation at this point. Recall that in order to get positive energy fluxes, we need to satisfy the following constraints for any CFT in 4 dimensions \cite{Hofman}:
\begin{equation}
1-\frac{t_2}{3}-\frac{2}{15}t_4\geq 0 \,,\quad1+\frac{t_2}{6}-\frac{2}{15}t_4\geq 0 \,, \quad 1+\frac{t_2}{3}+\frac{8}{15}t_4\geq 0\,,
\end{equation}
where $(c-a)/c=8\alpha/(1+8\alpha)=t_2/6+4 t_4/45$ for us. Remarkably, the ranges in \eqref{eqn45x8} satisfy all these three inequalities. In other words, our findings are within the theoretical limits set forth in \cite{Hofman}.  That this could happen was not at all guaranteed and we have examples of parameter spaces (e.g., considering the lower range for $M/\pi T$ to be 1.27) which would violate the above constraints.

Of course our numerical analysis should be taken with a grain of salt as we did not include the most general corrections possible. However, it will be a useful starting point for any future work on improving the AdS/CFT phenomenology program initiated in \cite{Buchel1,Buchel2}. For instance, it will be interesting to see the effect of adding a chemical potential to the corrections \cite{Myers1} or in the $\mathcal{N}=2^*$ model \cite{nstar}.

%
%

\vskip 2cm

\noindent {\bf Acknowledgments:} We thank Robert C. Myers for numerous
discussions and suggestions. Research at Perimeter Institute is supported by the
Government of Canada through Industry Canada and by the Province of
Ontario through the Ministry of Research \& Innovation. A.~Sinha thanks Perimeter Institute for hospitality during the course of this work.

\appendix

\section{Gauge conditions and other functions for eight derivative correction}
\label{appA}

For the eight derivative correction \eqref{eqn23}, the variables
mentioned in \eqref{eqn23x12} are:
\begin{align}
c_1(r) \,=\,&  -H_1(r) \frac{2 R_0^4 }{r (3 r^4 - R_0^4)}  - H_0(r) \frac{2 R_0^4}{L^6 r^{12} (3 r^4 - R_0^4)^3} (675 r^{20} + 648 L^4 M_0^2 r^{14} R_0^4 \notag \\
& - 225 r^{16} R_0^4 -1176 L^4 M_0^2 r^{10} R_0^8 - 2808 r^{12} R_0^8 + 632 L^4 M_0^2 r^6 R_0^{12}  \notag \\
& +876 r^8 R_0^{12} - 104 L^4 M_0^2 r^2 R_0^{16} + 1797 r^4 R_0^{16} -1275 R_0^{20}) + \frac{dH_0(r)}{dr} \times \notag \\
&  \frac{8 R_0^8 (r^4 - R_0^4)(2 L^4 M_0^2 r^2 (3 r^4 - R_0^4) - 3 (36 r^8 - 95 r^4 R_0^4 + 49 R_0^8))}{L^6 r^{11} (3 r^4 - R_0^4)^2}\,,
\label{eqn49} \\
d_1(r) \,=\,& \frac{r^2}{2L^3 M_0^2}\frac{d b_1(r)}{d r}-\frac{r^2R_0^{4}}{(3r^{4}-R_0^{4})L^3 M_0^2} \frac{d H_1(r)}{d r} -\frac{12 r^{4}R_0^{4}}{(3r^{4}-R_0^{4})^2 L^2 M_0^2}H_1(r) \notag \\
& + H_0(r) \frac{4 R_0^4}{L^3 M_0^3 r^{11} (-3 r^4 + R_0^4)^4}\Big(-2 L^8 M_0^5 r^4 R_0^4 (3 r^4 - R_0^4)^3  \notag \\
& \; + 6 L^6 M_1 r^{16} (-3 r^4 + R_0^4)^2 + L^4 M_0^3 r^2 R_0^4 (-1944 r^{16} + 927 r^{12} R_0^4  \notag \\
& \;+ 1041 r^8 R_0^8 -843 r^4 R_0^{12} + 155 R_0^{16}) - 9 M_0 (225 r^{24}  - 961 r^{16} R_0^8 \notag \\
& \; + 112 r^{12} R_0^{12} + 839 r^8 R_0^{16} -80 r^4 R_0^{20} - 135 R_0^{24})\Big) \notag \\
& - \frac{d H_0(r)}{d r} \frac{1}{2 L^9 M_0^3 r^{10} (3 r^4 - R_0^4)^3}\Big(3 L^6 M_1 r^{12} (3 r^4 - R_0^4)^2 (r^4 + R_0^4) \notag \\
& \; - 32 L^4 M_0^3 r^2 R_0^8 (9 r^{12} + 15 r^8 R_0^4 - 21 r^4 R_0^8 + 5 R_0^{12}) + 18 M_0 R_0^4 (75\, r^{20}  \notag \\
& \; + 551 r^{16} R_0^4 - 1376 r^{12} R_0^8 + 324 r^8 R_0^{12} + 741 r^4 R_0^{16} - 315 R_0^{20})\Big)\,.
\label{eqn50}
\end{align}

The final equation of motion for $H_1(r)$ is following:
\begin{align}
& r (r^4-R_0^4) \frac{d^2 H_1(r)}{d^2 r}+(5r^4-R_0^4)\frac{d H_1(r)}{d r}+\frac{r (64 r^2 R_0^8 + L^4 M_0^2  (-3 r^4 + R_0^4)^2)}{ (3 r^4 - R_0^4)^2}H_1(r)  \notag \\
& =\, \frac{H_0(r)}{L^6 r^{13} (3 r^4 - R_0^4)^4}\Big(-2 L^{10} M_0 M_1 r^{14} (3 r^4 - R_0^4)^4 -48 L^8 M_0^4 r^4 R_0^8 (3 r^4 - R_0^4)^4 \notag \\
& \quad \; - 192 R_0^8 \big(450 r^{24} + 1803 r^{20} R_0^4 - 9249 r^{16} R_0^8 + 8288 r^{12} R_0^{12} + 522 r^8 R_0^{16}  \notag \\
& \quad \; - 2219 r^4 R_0^{20}  + 405 R_0^{24}\big)- L^4 M_0^2 r^2 R_0^4 (6075 r^{24} - 320841 r^{20} R_0^4 + 952785 r^{16} R_0^8  \notag \\
& \quad \; - 1022070 r^{12} R_0^{12} + 507657 r^8 R_0^{16} - 117881 r^4 R_0^{20} + 10291 R_0^{24})\Big) \notag \\
&\quad -\frac{d H_0(r)}{d r} \frac{4 R_0^4 (-r^4 + R_0^4)}{ r^{12} (3 r^4 - R_0^4)^3}  \Big(-2025 r^{20} - 4680 L^4 M_0^2 r^{14} R_0^4 + 51111 r^{16} R_0^4  \notag \\
& \quad \; + 6024 L^4 M_0^2 r^{10} R_0^8 - 169776 r^{12} R_0^8 - 2328 L^4 M_0^2 r^6 R_0^{12} + 209412 r^8 R_0^{12} \notag \\
& \quad  \;+ 280 L^4 M_0^2 r^2 R_0^{16} - 94623 r^4 R_0^{16} + 12573 R_0^{20}\Big)\,,
\label{eqn51}
\end{align}
and the solution near the horizon is given by
\begin{align}
H_1(r) =& -\frac{M_0 L^3 (48 L^4 M_0^3 + 1001 M_0 R_0^2 + 2  M_1 R_0^2) (L^5 M_0^2 + R_0^2 (16 L + (16 + L^2 M_0^2) R_0))}{R_0^2(L^4 M_0^2  + 16 R_0^2)^2} \notag \\
& + (r-R_0)\frac{M_0 L^3 (16 + M_0^2 L^2) (48 L^4 M_0^3 + 1001 M_0 R_0^2 + 2 M_1 R_0^2)}{4 R_0^2 (L^4 M_0^2 + 16 R_0^2)}\,.
\label{eqn23x16}
\end{align}
The equation of motion for $H_{v1}(r)$ is following:
\begin{align}
&r \frac{d^2H_{v1}(r)}{dr^2}+5\frac{dH_{v1}(r)}{dr}+\frac{L^4 M_{v0}^2 r}{r^4-R_0^4} H_{v1}(r)= - \frac{M_{v0}H_{v0}(r)}{L^2 r^{11} (r^4 - R_0^4)}\Big(2 L^6 M_{v1} r^{12}  \notag \\
& \qquad + 48 L^4 M_{v0}^3 r^2 R_0^8 + 15 M_{v0} R_0^4 (5 r^8 - 91 r^4 R_0^4 + 85 R_0^8)\Big) \notag \\
& \qquad - \frac{dH_{v0}(r)}{dr}\frac{80 R_0^8}{L^6 r^{12}}\big(8 L^4 M_0^2 r^2 - 36 r^4 + 27 R_0^4\big)\,,
\label{eqn51x1}
\end{align}
and the solution near the horizon is given by
\begin{equation}
H_{v1}(r)=- (r-R_0)\frac{48 L^8 M_0^4 + 2545 L^4 M_0^2 R_0^2 + 2 L^{4} M_0 M_1 R_0^2 - 2880 R_0^4}{L R_0^2 (L^4 M_0^2 + 20 R_0^2)}\,.
\label{eqn24x4}
\end{equation}
The equation of motion for $H_{s1}(r)$ is given by
\begin{align}
&r (r^4 - R_0^4)\frac{d^2H_{s1}(r)}{dr^2}+(5r^4-R_0^4)\frac{dH_{s1}(r)}{dr}+r L^4 M_{s0}^2  H_{s1}(r)\notag \\
&=-\frac{M_{s0} H_{s0}(r)}{L^2 r^{11}}\Big(2 L^6 M_{s1} r^{12} + 48 L^4 M_{s0}^3 r^2 R_0^8 + M_{s0} R_0^4 (75 r^8 - 789 r^4 R_0^4 + 851 R_0^8)\Big)\notag \\
& \quad  -\frac{d H_{s0}(r)}{dr} \frac{12 R_0^4 (r^4 - R_0^4)}{L^6 r^{12}}\big(25 r^8 + 40 L^4 M_{s0}^2 r^2 R_0^4 - 94 r^4 R_0^4 - 129 R_0^8\big)\,,
\label{eqn51x2}
\end{align}
and the solution near the horizon is
\begin{align}
H_{s1}(r)=& -\frac{2 (48 L^8 M_{s0}^5 + 2441 L^4 M_{s0}^3 R_0^2 + 2 L^{10} M_{s0}^2 M_{s1} R_0^2 +
4744 M_{s0} R_0^4 + 16 L^6 M_{s1} R_0^4)}{L^6 M_{s0} R_0^2 (L^4 M_{s0}^2 + 16 R_0^2)} \notag \\
&+(r-R_0)\frac{M_{s0}^2 (48 L^8 M_{s0}^4 + 3977 L^4 M_{s0}^2 R_0^2 + 2 L^{10} M_{s0} M_{s1} R_0^2 + 7296 R_0^4)}{4 L^2 R_0^5 (L^4 M_{s0}^2 + 16 R_0^2)}\,.
\label{eqn24x9}
\end{align}
We display below the corrections to the geometry for the various higher derivative terms considered in this paper
\begin{table}[ht]
\centering
\begin{tabular}{ | c | c | c |   }
  \hline
  Term & $f_1(r)$ & $g_1(r)$   \\
  &  &  \\
  \hline
  $W^2$ & $\displaystyle -\frac{2 R_0^4}{r^4} + \frac{8 (50 r^4 R_0^4 - 17 R_0^8)}{3 r^8} \alpha$ & $\displaystyle -\frac{2 R_0^4}{r^4}+ \frac{8 (50 r^4 R_0^4 - 87 R_0^8)}{3 r^8} \alpha$ \\
  \hline
  $W^3$ & $\displaystyle \frac{-14 r^4 R_0^4 + 6 R_0^8}{ r^8}$ & $\displaystyle \frac{2 (-7 r^4 R_0^4 + 13 R_0^8)}{ r^8}$  \\
  \hline
  $W^4_1$  & $\displaystyle \frac{28 R_0^4 (5 r^8 + 5 r^4 R_0^4 - 3 R_0^8)}{r^{12}}$ & $\displaystyle \frac{28 R_0^4 (5 r^8 + 5 r^4 R_0^4 - 19 R_0^8)}{r^{12}}$  \\
  \hline
  $W^4_2$   & $\displaystyle -\frac{56 R_0^4 (5 r^8 + 5 r^4 R_0^4 - 3 R_0^8)}{ r^{12}}$ & $\displaystyle -\frac{56 R_0^4 (5 r^8 + 5 r^4 R_0^4 - 19 R_0^8)}{ r^{12}}$  \\
  \hline
  $W^4_3$   & $\displaystyle -\frac{15 R_0^4 (5 r^8 + 5 r^4 R_0^4 - 3 R_0^8)}{ r^{12}}$ & $\displaystyle -\frac{15 R_0^4 (5 r^8 + 5 r^4 R_0^4 - 19 R_0^8)}{ r^{12}}$  \\
  \hline
  $W^4_4$  & $\displaystyle -\frac{29 R_0^4 (5 r^8 + 5 r^4 R_0^4 - 3 R_0^8)}{ r^{12}}$ & $\displaystyle -\frac{29 R_0^4 (5 r^8 + 5 r^4 R_0^4 - 19 R_0^8)}{ r^{12}}$  \\
  \hline
  $W^4_5$  & $\displaystyle -\frac{74 R_0^4 (5 r^8 + 5 r^4 R_0^4 - 3 R_0^8)}{ r^{12}}$ & $\displaystyle -\frac{74 R_0^4 (5 r^8 + 5 r^4 R_0^4 - 19 R_0^8)}{ r^{12}}$  \\
  \hline
\end{tabular}
\caption{Corrections to metric}
\label{table2}
\end{table}

\section{Higher derivative corrections to $\eta/s$} \label{transport}

In this appendix, we review the calculation of the shear viscosity of the holographic
supersymmetric plasma with eight-derivative correction \eqref{eqn23}. We use the
Kubo formula that relates the transport coefficients of the plasma to
the field theory correlators and these correlators can be calculated
using holographic techniques \cite{Policastro}. For higher
derivative corrections we follow \cite{Buchel, Myers1}. 

Following the prescription set in \cite{Myers1}, we use $u=R_0^2/r^2$ which is more convenient for the hydrodynamic
calculations. The horizon is at $u=1$
and the black hole solution for \eqref{eqn22} is given by
\begin{equation}
ds^2=-\frac{R_0^2}{L^2}\frac{f(u)}{u}dt^2+\frac{L^2}{4u^2g(u)}du^2+\frac{R_0^2}{L^2}\frac{1}{u}(dx^2+dy^2+d\tau^2)\,,
\label{eqn31}
\end{equation}
where
\begin{align}
f(u)\quad =& \quad f_0(u)(1+\gamma_1 f_1(u))\,, \notag \\
g(u)\quad =& \quad f_0(u)(1+\gamma_1 g_1(u)) \notag \\
\textrm{and}\quad f_0(r)\quad =& \quad (1-u^2)\,, \notag \\
f_1(u)\quad =& \quad -15 u^2 (-5 - 5 u^2 + 3 u^4)\,, \notag \\
g_1(u)\quad =& \quad -15 u^2 (-5 - 5 u^2 + 19 u^4)\,.
\label{eqna25}
\end{align}
Kubo formula relates the shear viscosity to the low frequency and
zero momentum limit of the retarded Green's function of the stress
tensor
\begin{equation}
\eta=-\lim_{\omega \rightarrow 0}\frac{1}{\omega} \textrm{Im} G_{xy,xy}^{R}(\omega, k=0)\,,
\label{eqn33}
\end{equation}
where
\begin{equation}
G_{xy,xy}^{R}(\omega, k=0)=-i\int dt\, dx\, dy\, d\tau\, e^{i\omega t}\langle[T_{xy}(x),T_{xy}(0)]\rangle\,.
\label{eqn34}
\end{equation}
Now we translate the calculation of the correlator to the dual gravity
by first calculating the effective action for the metric perturbation
$h_x^{\; y}(t,u)=\int \frac{d^4k}{(2\pi)^4}\phi_k(u)e^{-i\omega t+i k
\tau}$. Evaluating the action \eqref{eqn22} to quadratic order in the
fluctuation $\phi_k(u)$ yields
\begin{align}
I^{(2)}_{\phi}=\frac{1}{2\ell_p^3}\int & \frac{d^4k}{(2\pi)^4} du (A(u,\omega)\phi''_k \phi_{-k}+B(u,\omega)\phi'_k\phi'_{-k}+ C(u,\omega)\phi'_k\phi_{-k} \notag \\
& +D(u,\omega)\phi_k\phi_{-k} +E(u)\phi''_k\phi''_{-k}+F(u)\phi''_k\phi'_{-k}) + \mathcal{K} \,,
\label{eqn35}
\end{align}
where $\mathcal{K}$ is the generalized Gibbons-Hawking boundary term
and its detailed expression can be found in \cite{Myers1}. Now we
follow the arguments given in \cite{Myers1} and directly read off the
shear viscosity from the action \eqref{eqn35}
\begin{align}
\eta=\frac{1}{\ell_p^3}(\kappa_1(u)+\kappa_2(u))_{u=1}\,,
\label{eqn36}
\end{align}
where
\begin{align}
\kappa_1(u)=& \lim_{\omega\to 0}\sqrt{-\frac{g_{uu}(u)}{g_{tt}(u)}}\left(A(u,\omega)-B(u,\omega)+\frac{F'(u,\omega)}{2}\right)\,, \notag \\
\kappa_2(u)=& \lim_{\omega\to 0} \left( E(u,\omega)\left( \sqrt{-\frac{g_{uu}(u)}{g_{tt}(u)}}\right)' \right)'\,.
\label{eqn37}
\end{align}
Note that in \eqref{eqn36} we have evaluated the quantities at the
horizon whereas in \eqref{eqn33} it was evaluated at the asymptotic
boundary (at $u=0$). For details of the arguments about these
calculations, the reader is suggested to refer to \cite{Myers1}.
The functions $A,B,C,D,E,F$ at the horizon are given by
\begin{align}
A(u,0)\quad =& \quad \frac{2 R_0^4 (1 - u^2)}{L^5 u}-\frac{10 R_0^4 u (1 - u^2) (15 + 15 u^2 - 29 u^4) \gamma}{ L^{5} }\,, \notag \\
B(u,0)\quad =& \quad \frac{3 R_0^4 (1 - u^2)}{2 L^5 u} \notag \\
& \qquad + \frac{R_0^4 u (-45 (5 - 16 u^4 + 11 u^6) + L^6 (8 u^2 - 52 u^4 + 76 u^6)) \gamma}{2 L^{5}}\,, \notag \\
E(u,0)\quad =& \quad \frac{16 R_0^4 u^5 (1 - u^2)^2  \gamma}{L^3} \,,\notag \\
F(u,0)\quad =& \quad \frac{16 R_0^4 u^4 (1 - 3 u^2 + 2 u^4) \gamma}{L^4}\,.
\label{eqn52}
\end{align}
Using these we find that the shear
viscosity is
\begin{align}
\eta=&\frac{\pi^3 L^3 T^3}{2 \ell_p^3}(1+135\gamma)\,.
\label{eqn44}
\end{align}
The ratio of shear viscosity with entropy density \eqref{eqn23x8} is
\begin{align}
\frac{\eta}{s}=&\frac{1}{4\pi}(1+120 \gamma)\,.
\label{eqn45}
\end{align}

\begin{table}[ht]
\centering
\begin{tabular}{ | c | c | c | c | c | }
  \hline
  Term  & $\left( \frac{\pi L^2}{R_0} \right)\times$ Temperature & $\left(\frac{2 l_p^3}{3 \pi^4 L^3 T^4}\right)\times$ Free energy & $\left( \frac{2 l_p^3}{\pi^3 L^3 T^3}\right) \times \eta$ \\
  \hline
  $W^2$ & $\displaystyle \left( 1-2\alpha-\frac{16}{3} \alpha^2 \right)$ & $ ( 1+18\alpha + 24 \alpha^2)$ & $(1+10\alpha+136 \alpha^2)$  \\
  \hline
  $W^3$ & $(1 + 2 \beta)$ &$(1+6 \beta)$ &  $(1-90\beta)$  \\
  \hline
  $W^4_1$ & $(1 - 28 \gamma_1)$ &$(1-28 \gamma_1)$ & $(1-444\gamma_1)$   \\
  \hline
  $W^4_2$ & $(1 + 56 \gamma_2)$ & $(1+56 \gamma_2)$ & $(1 + 888 \gamma_2)$  \\
  \hline
  $W^4_3$ & $(1 + 15 \gamma_3)$ &$(1+15 \gamma_3)$ & $(1 + 135 \gamma_3)$   \\
  \hline
  $W^4_4$ & $(1 + 29 \gamma_4)$ &$(1+29 \gamma_4)$ & $(1 + 357 \gamma_4)$   \\
  \hline
  $W^4_5$  & $(1 + 74 \gamma_5)$ &$(1+ 74 \gamma_5)$ & $(1 + 762 \gamma_5)$  \\
  \hline
\end{tabular}
\caption{Corrections to thermodynamics and shear viscosity}
\label{table3}
\end{table}

 The results for other correction
terms are stated in Table \ref{table5} and \ref{table3}. For the sake
of completeness of the discussion, we have borrowed the results from
\cite{Banerjee} and \cite{Buchel2} for the second order corrections to
$\eta$.

\end{document}